\pdfoutput=1

\documentclass[numberedappendix,apj]{emulateapj}
\usepackage{apjfonts}
\usepackage[colorlinks=true,urlcolor=blue,linkcolor=blue,citecolor=blue]{hyperref}
\usepackage{amsmath}
\usepackage{natbib}
\usepackage{graphicx}

\usepackage{etoolbox}
\makeatletter
\patchcmd{\NAT@citex}
  {\@citea\NAT@hyper@{\NAT@nmfmt{\NAT@nm}\NAT@date}}
  {\@citea\NAT@nmfmt{\NAT@nm}\NAT@hyper@{\NAT@date}}
  {}
  {}
\patchcmd{\NAT@citex}
  {\@citea\NAT@hyper@{%
     \NAT@nmfmt{\NAT@nm}%
     \hyper@natlinkbreak{\NAT@aysep\NAT@spacechar}{\@citeb\@extra@b@citeb}%
     \NAT@date}}
  {\@citea\NAT@nmfmt{\NAT@nm}%
   \NAT@aysep\NAT@spacechar%
   \NAT@hyper@{\NAT@date}}
  {}
  {}
\patchcmd{\NAT@citex}
  {\@citea\NAT@hyper@{%
     \NAT@nmfmt{\NAT@nm}%
     \hyper@natlinkbreak{\NAT@spacechar\NAT@@open\if*#1*\else#1\NAT@spacechar\fi}%
       {\@citeb\@extra@b@citeb}%
     \NAT@date}}
  {\@citea\NAT@nmfmt{\NAT@nm}%
   \NAT@spacechar\NAT@@open\if*#1*\else#1\NAT@spacechar\fi%
   \NAT@hyper@{\NAT@date}}
  {}
  {}
\makeatother

\graphicspath{{./imgs/}}

\def\epsff{\epsilon_{\rm ff}}
\def\epssf{\epsilon_{\rm sf}}
\def\epsw{\epsilon_{\rm w}}
\def\avir{\alpha_{\rm vir}}
\def\rhosf{{\rho}_{\rm sf}}
\def\rhosfr{\dot{\rho}_\star}
\def\Ssfr{\dot\Sigma_\star}
\def\Sgas{\Sigma_{\rm g}}
\def\fH2{f_{\rm H_2}}
\def\vrms{\sigma}
\def\cs{c_{\rm s}}
\def\tff{t_{\rm ff}}
\def\tsf{t_{\rm sf}}
\def\tcr{t_{\rm cr}}
\def\tdep{t_{\rm dep}}
\def\tarm{t_{\rm arm}}
\def\tdec{t_{\rm dec}}
\def\warm{w_{\rm arm}}
\def\varm{v_{\rm arm}}

\def\pc{\mathrm{~pc}}
\def\kpc{\mathrm{~kpc}}
\def\Myr{\mathrm{~Myr}}
\def\Gyr{\mathrm{~Gyr}}
\def\kms{\mathrm{~km~s^{-1}}}
\def\cc{\mathrm{~cm^{-3}}}
\def\Msun{\mathrm{~M_{\odot}}}
\def\Msunyr{\mathrm{~M_{\odot}~yr^{-1}}}
\def\Msunpc2{\mathrm{~M_{\odot}~pc^{-2}}}
\def\Msunyrkpc2{\mathrm{~M_{\odot}~yr^{-1}~kpc^{-2}}}
\def\K{\mathrm{~K}}
\def\dex{\mathrm{~dex}}

\begin{document}


\shorttitle{Non-universal star formation efficiency in turbulent ISM}
\shortauthors{Semenov, Kravtsov, Gnedin}
\slugcomment{{Accepted for publication in the Astrophysical Journal:} April 27, 2016} 

\title{Non-universal star formation efficiency in turbulent ISM}

\author{Vadim A. Semenov\altaffilmark{1,2,$\star$}, Andrey V. Kravtsov\altaffilmark{1,2,3} and Nickolay Y. Gnedin\altaffilmark{1,2,4}}


\altaffiltext{1}{Department of Astronomy \& Astrophysics, The University of Chicago, Chicago, IL 60637 USA}
\altaffiltext{2}{Kavli Institute for Cosmological Physics, The University of Chicago, Chicago, IL 60637 USA}
\altaffiltext{3}{Enrico Fermi Institute, The University of Chicago, Chicago, IL 60637 USA}
\altaffiltext{4}{Fermilab Center for Particle Astrophysics, Fermi National Accelerator Laboratory, Batavia, IL 60510-0500 USA}
\altaffiltext{$\star$}{semenov@uchicago.edu}


\begin{abstract}

We present a study of a star formation prescription in which star formation efficiency depends on local gas density and turbulent velocity dispersion, as suggested by direct simulations of SF in turbulent giant molecular clouds (GMCs). We test the model using a simulation of an isolated Milky Way-sized galaxy with a self-consistent treatment of turbulence on unresolved scales. We show that this prescription predicts a wide variation of local star formation efficiency per free-fall time, $\epsff \sim 0.1 - 10\%$, and gas depletion time, $\tdep \sim 0.1 - 10 \Gyr$. In addition, it predicts an effective density threshold for star formation due to suppression of $\epsff$ in warm diffuse gas stabilized by thermal pressure. We show that the model predicts star formation rates in agreement with observations from the scales of individual star-forming regions to the kiloparsec scales. This agreement is non-trivial, as the model was not tuned in any way and the predicted star formation rates on all scales are determined by the distribution of the GMC-scale densities and turbulent velocities $\vrms$ in the cold gas within the galaxy, which is shaped by galactic dynamics. The broad agreement of the star formation prescription calibrated in the GMC-scale simulations with observations, both gives credence to such simulations and promises to put star formation modeling in galaxy formation simulations on a much firmer theoretical footing.

\end{abstract}

\section{Introduction}
\label{sec:introduction}
\setcounter{footnote}{0}

Numerical simulations of galaxy formation require modelling of how gas is converted into stars, the process that is not yet completely understood. Star formation is thus implemented in simulations using empirically motivated, phenomenological prescriptions. For example, in almost all such simulations star particles are formed from gas that satisfies a set of fairly ad hoc criteria, such as requiring that the gas density exceeds a certain threshold, $\rho > \rhosf$, or that the gas is locally compressed. Star formation rate (SFR) in this gas is usually parametrized as:
\begin{equation}
\label{eq:SFR}
\rhosfr = \epssf \frac{\rho}{\tsf},
\end{equation}
where $\epssf$ is an efficiency at which gas is converted into stars over the time scale $\tsf$. 

This basic approach has not changed since the early 1990s, when it was introduced \citep[][]{Katz.1992,Cen.Ostriker.1992}, with the exception of SFR modulation with local molecular hydrogen fraction \citep{Robertson.Kravtsov.2008,Gnedin.etal.2009} and an attempt to account for turbulent gas motions in the identification of star-forming regions by \citet{Hopkins.etal.2013}. The choices for the star formation eligibility criteria and values of $\epssf$ and $\tsf$ do vary significantly from study to study. The most common recent choice is to adopt $\tsf=\tff$, where the free-fall time is defined as $\tff = \sqrt{3\pi/32G\rho}$. In this case, $\epssf$ has a meaning of star formation efficiency (SFE) per free-fall time, $\epsff$. 

In some studies specific values of $\epsff$ are chosen to reproduce the observed relation between gas surface density and star formation rate \citep{Schmidt.1959,Schmidt.1963,Kennicutt.1998,Bigiel.etal.2008,Kennicutt.Evans.2012}. In other studies, the assumed $\epsff$ value is based on the independent reasoning of what it should be in real star-forming regions. Such simulations require self-regulation by a complex interplay between star formation and feedback to produce the observed SFE on the scales of entire galaxy \citep[e.g.,][]{Stinson.etal.2006,Ceverino.Klypin.2009,Hopkins.etal.2011}. In general, recent results indicate that when effects of feedback are weak, the star formation on the galactic scale is highly sensitive to the assumed input value of $\epsff$, while when feedback is strong the sensitivity is significantly weaker \citep[e.g.,][]{Agertz.etal.2013}.

Regardless of the $\epsff$ value choice, almost all galaxy formation simulations assume that this value is universal in space and time. Consequences of such assumption on galaxy evolution remain unexplored. 

There is a growing number of observational indications that $\epsff$ in star-forming giant molecular clouds (GMCs) varies widely from $\epsff \sim 0.1\%$ to $\sim 30\%$ \citep{Heiderman.etal.2010,Lada.etal.2010,Murray.2011,Evans.etal.2014}. There is also evidence that in the Milky Way the overall molecular gas forms stars at a considerably lower efficiency than typical well-studied dense star-forming regions. In particular, the molecular gas depletion time for the entire Galaxy is $t_{\rm dep, MW}\equiv M_{\rm H_2,MW}/\dot{M}_{\star,\rm MW} \sim 10^9\, \Msun/(1.5\, \Msunyr) \sim 0.7\ \Gyr$. This is an order of magnitude longer time scale than the typically inferred in active star-forming molecular clouds: $t_{\rm dep, GMC}=t_{\rm ff}/\epsff\sim 5\Myr /0.05\sim 100\Myr$. 
 
The strong variation of $\epsff$ also naturally arises in models of star formation in turbulent medium. These models are motivated by the observed scaling relations between size, mass and velocity dispersion of GMCs \citep{Larson.1981} that are consistent with a supersonic turbulent cascade in interstellar medium (ISM) \citep[e.g.,][]{Kritsuk.etal.2013}. Over the past two decades several analytical models for star formation in turbulent GMCs have been proposed \citep{Krumholz.McKee.2005,Padoan.Nordlund.2011,Federrath.Klessen.2012,Hennebelle.Chabrier.2013}. Star formation in such models is assumed to occur in the densest, gravitationally unstable tail of the gas density distribution. The SFR then is estimated assuming {\it static} gas density probability distribution function (PDF), predicted by turbulent cloud simulations. The dependence of star formation on properties of turbulence, such as the average Mach number, arises in these models via dependence of the gas density PDF on such properties.  As a result, these models generically predict variation of SFE with the Mach number, magnetic fields strength and the so-called virial parameter $\avir$, that parametrizes relative importance of turbulence with respect to gravity in a given region. 

Recent high-resolution simulations of self-gravitating turbulent clouds have reached the stage of resolving gravitationally unstable dense clumps, thus allowing modelling of star formation in GMCs \citep{Clark.etal.2005,Price.Bate.2009,Wang.etal.2010,Krumholz.2012,Padoan.etal.2012,Federrath.2015}. The density PDF and other relevant physical properties and processes are evolved more or less self-consistently and SFE is usually estimated directly by counting the total mass of sink particles formed over a certain period of time. Remarkably, \citet{Padoan.etal.2012} found that in such simulations the relatively complex dependencies of star formation on turbulence properties suggested by the analytic models are replaced by a simple exponential relation between $\epsff$ and cloud virial parameter, $\avir$, only. 

These results pave a way to modelling $\epsff$ in galaxy formation simulations. However, implementation of such model requires  information on gas motions on sub-GMCs scale to compute $\avir$, which is not readily available. This is because star formation is implemented on the resolution scale of such simulations, on which gas motions are strongly affected by numerical viscosity. For example, in grid-based hydrodynamics codes this effect becomes significant on scales as large as $\approx 10-20$ computational cells \citep{Kritsuk.etal.2011} and should be even larger in simulations using smooth particle hydrodynamics due to explicit artificial viscosity. Thus, even though state-of-the-art galaxy formation simulations reach resolution from few to hundreds of parsec, gas motions can only be trusted on scales at least an order of magnitude larger. 

One of the promising alternative ways to assess small-scale turbulence in hydrodynamical simulations is subgrid-scale (SGS) modelling. Models of this kind have been developed and extensively used in simulations of terrestrial turbulent subsonic and supersonic flows \citep[see e.g.,][for review]{Sagaut,Garnier.etal}. The main idea of these models lies in averaging hydrodynamics equations of turbulent flow on a certain scale $\Delta$, that is often associated with a computational cell size. The resulting set of equations then governs flows on scales $>\Delta$ and contains terms that depend on integral properties of turbulence at scales $<\Delta$. Turbulent energy at scales $<\Delta$ is also modelled as a separate field similar to gas thermal energy. 

SGS turbulence modelling was implemented in a number astrophysical studies \citep[e.g.,][]{Ropke.etal.2007,Maier.etal.2009,Iapichino.etal.2011}. \citet{Schmidt.etal.2014} describe such SGS model in the context of cosmological structure formation. Notably, galaxy formation simulations have already started using this model for non-resolved turbulence treatment as well \citep{Latif.etal.2013,Braun.etal.2014,Braun.Schmidt.2015}. 

In this study, we use the SGS turbulence model of \citet{Schmidt.etal.2014} to investigate the variation of SFE in an isolated disk simulation of a Milky Way-sized galaxy. The main goal of this paper is to assess the viability of using the GMC-scale modelling results of \citet{Padoan.etal.2012} in galaxy-scale simulations. 

Our results show that such model predicts a wide variation of $\epsff$ and is broadly consistent with observations both on the GMC scale and on the scale of the entire galaxy. Furthermore, we demonstrate that this model naturally results in an effective physical density threshold for star formation. This allows us to avoid arbitrary assumptions about such critical density and temperature for star formation, often adopted in modern galaxy formation simulations. The model also does not require explicit modelling of molecular hydrogen. The SF prescription based on the \citet{Padoan.etal.2012} results, thus, looks promising for galaxy formation simulations.

The rest of the paper is organized as follows. In Section~\ref{sec:method} we outline our methods and initial conditions, discuss the SGS turbulence model and the adopted prescriptions for star formation and feedback. In Section~\ref{sec:results} we present our main results on the global disk structure, properties of the SGS turbulence and the resulting SFRs. We also show that the predicted wide distribution of SFE is consistent with local observations of GMCs as well as with large-scale SFRs of Milky Way and nearby galaxies. In Section~\ref{sec:discussion} we discuss our main results and compare them to previous numerical findings. Section~\ref{sec:summary} summarizes our results and conclusions.

\section{Numerical model}
\label{sec:method}

In order to investigate the effect of small-scale turbulence on star formation we simulate an isolated Milky Way-sized disk galaxy with a model for SGS turbulence, star formation prescription motivated by GMC-scale simulations, and stellar feedback model based on subgrid evolution of supernova remnants.

We perform our simulation with the Adaptive Refinement Tree (ART) $N$-body and gas dynamics code \citep{Kravtsov.1999,Kravtsov.etal.2002,Rudd.etal.2008}. The ART code is a Eulerian code that employs Adaptive Mesh Refinement (AMR) technique with the Fully Threaded Tree data structure \citep{Khokhlov.1998} and a shock-capturing second-order Godunov-type method \citep{Colella.Glaz.1985} with piecewise linear reconstruction \citep{vanLeer.1979} to compute hydrodynamical fluxes. The Poisson equation for gas self-gravity and collisionless dynamics of stellar and dark matter particles is solved using the FFT on the zeroth uniform grid level and relaxation method on the refinement levels \citep{Kravtsov.etal.1997}. Gravitational potential and accelerations are used to update positions and velocities of collisionless particles and are also applied in the gas momentum and energy equations as source terms.

The SGS turbulence model is implemented by introducing additional terms, that mediate interactions of resolved and unresolved flows, into momentum and energy equations (see Section~\ref{sec:method:SGST} for details). Our code separately follows the evolution of thermal and subgrid turbulent energies. Thermal energy is used to follow dissipative thermodynamics of the gas, while the subgrid turbulent energy is used to compute star formation efficiency, as well as the SGS terms in the fluid dynamics equations. In order to prevent strong fragmentation of cold gas on resolution scale we add artificial pressure support \citep[e.g.,][]{Machacek.etal.2001}.

\subsection{Subgrid-scale turbulence}
\label{sec:method:SGST}

To model SGS turbulence we employ the scale separation technique, where a large-scale flow is governed by filtered hydrodynamical equations, whereas small-scale motions are described by an additional hydrodynamical field. In our simulation we use the SGS model described by \citet{Schmidt.etal.2014} for application in cosmological AMR simulations. Here we briefly outline the main components and properties of this model. The more extensive description can be found in the original paper by \citet{Schmidt.etal.2014}.

Model equations follow from applying a spatial filter of scale $\Delta$, which we take to correspond to the grid cell scale, to ordinary hydrodynamical equations. The resulting set of equations governs flow on resolved scales ($>\Delta$) and contains additional terms and a new equation for SGS turbulent energy density $K$:
\begin{equation}
\label{eq:K}
\frac{\partial}{\partial {t}} {K} + {\nabla}_i ( {u}_i {K} ) = - P_K {\nabla}_i {u}_i - {\varepsilon}  + {\tau}_{ij} {\nabla}_i {u}_j + {\nabla}_i F_{i} + S_{\rm SN},
\end{equation}
where ${u}_i$ is resolved gas velocity, $P_K = 2K/3$ is turbulent pressure, $\varepsilon$ is a rate of decay into thermal energy, ${\tau}_{ij} {\nabla}_i {u}_j$ is viscous production by cascade from resolved scales, ${\nabla}_i F_{i}$ is turbulent diffusion and $S_{\rm SN}$ is sourcing by supernovae (see Section~\ref{sec:method:feedback} for the description of $S_{\rm SN}$). The total velocity dispersion of gas motions on unresolved scales is then derived from $K$ as $\vrms = \sqrt{2K/\rho}$.

Note that the SGS turbulent energy is very similar to thermal energy, as the latter results from integrating particles kinetic energies over velocity space in the derivation of hydrodynamical equations from the Boltzmann equation. For instance, the first term on the right-hand side of Equation~(\ref{eq:K}) is equivalent to the $PdV$-term in the equation for thermal energy. This term implies that as gas contracts (expands) $P_K$ does work and turbulent energy increases (decreases) similarly to thermal energy \citep{Robertson.Goldreich.2012}. Change of $K$ in this process depends on local compression rate ($-\nabla_i u_i$).

Likewise, the $\varepsilon$ and ${\tau}_{ij} {\nabla}_i {u}_j + S_{\rm SN}$ are equivalent to the cooling and heating terms in the thermal energy equation. We follow \citet{Schmidt.etal.2014} and assume exponential decay of $K$ into thermal energy over the time scale close to the turbulent cell crossing time, $\tdec \sim \Delta/\vrms$. Numerical studies of decaying MHD turbulence generally confirm fast dissipation over crossing time both in subsonic and supersonic regimes \citep[e.g.,][]{Gammie.Ostriker.1996,MacLow.etal.1998,Stone.etal.1998,Kim.Basu.2013}. 

Equations for resolved gas momentum and energy also include terms related to non-thermal pressure ($P_K$), turbulent viscosity (${\tau}_{ij}$) and diffusion (similar to ${\nabla}_i F_{i}$). The latter two terms are analogous to molecular viscosity and thermal conduction that appear in the hydrodynamical equations when different moments of the Boltzmann equation are integrated over velocity space.

The equations of viscous hydrodynamics and SGS turbulence both require closure relations for these transport terms in order to become solvable. In both cases these closure relations cannot be derived from the first principles and are chosen empirically. One of the common choices for the SGS turbulence is to adopt the closure relations similar to those used for usual viscosity and thermal conduction. Physically, this approach assumes that energy and momentum are transported on the filtering scale $\Delta$ mainly by the eddies of size $\Delta$, i.e. the largest unresolved eddies. Models that employ this assumption are known as the Large-Eddy Simulations (LES) and are widely used for simulations of both incompressible \citep{Sagaut} and supersonic \citep{Garnier.etal} turbulent flows \citep[see also][for a recent overview in the astrophysical context]{Schmidt.2014.book}.

In our simulation for the turbulent stress tensor ${\tau}_{ij}$ we use the large-eddy viscosity closure, given by Equation~(8) of \citet{Schmidt.Federrath.2011} with $C_2 = 0$ and $C_1 = 0.095$, appropriate for sub- and transonic regime. Our choice is justified by the fact that viscous production of turbulence in our simulated disk is important mainly in warm diffuse gas where SGS turbulence is subsonic and gas is only weakly compressible (see Section~\ref{sec:results:turbulence}). We checked that our implementation of the SGS model with such closure reproduces the correct distribution of $K$ in a low-resolution isotropic developed turbulence box simulation, when compared to a high-resolution direct simulation.

In the adopted closure, ${\tau}_{ij}$ depends on the local gradients of the resolved velocity field, and these gradients are interpreted as an onset of turbulent cascade on scale $\Delta$. Therefore, in this model turbulence can be artificially produced by large-scale velocity gradients (e.g., differential rotation, disk/halo interface, etc.). In order to suppress this spurious production \citet{Schmidt.etal.2014} suggest temporal averaging of simulated flow, so that ${\tau}_{ij}$ depends only on the gradients of fluctuating velocity part, in the so-called ``shear improved'' closure, first introduced by \citet{Leveque.etal.2007}. In our simulation we adopt exponential temporal filtering with a time window $t_{\rm si}=10\Myr$, i.e. turbulent energy is produced by cascade from velocity perturbations that develop faster than $t_{\rm si}$. We choose the value $t_{\rm si}=10\Myr$ to filter out the differential rotation, on the one hand, and to capture various developing disk instabilities, on the other hand. We checked that our results are not sensitive to a change of $t_{\rm si}$ by a factor of 2.

Although the SGS turbulence has a number of parameters, as described above, these parameters are calibrated using turbulence simulations and are not varied in galaxy formation simulations. In this sense, they do not really add tunable free parameters in such simulations. They do affect the solution, however, as a particular choice of the closure relation forms and their parameters controls all interactions between resolved and unresolved scales and may depend on flow configuration and turbulent Mach number. Generally, this might be considered as an important limitation of our model, as the specific closure adopted for our simulation was calibrated to reproduce the results of high-resolution simulations of developed isotropic transonic turbulence, while we apply it to a sheared gas flow in a stratified disk. However, we argue that this approach is still viable for prediction of the turbulent velocities in cold star-forming gas. Specifically, we checked that the resulting distribution of turbulent energy \textit{in cold gas} is not sensitive to a particular choice of ${\tau}_{ij}$ parametrization. This is because the turbulent energy in this gas is mostly determined by the interplay between heating by compression and viscous dissipation into heat (see Section~\ref{sec:results:turbulence}). Both these effects are insensitive to turbulent Mach number as indicated by numerical simulations of developed turbulence \citep{MacLow.etal.1998,Robertson.Goldreich.2012}.

One important limitation of our model is an assumption that unresolved turbulence on scale $\Delta$ is in the inertial regime. This assumption is made implicitly, because the direct simulations of developed turbulence, which were used to calibrate this model, do resolve the inertial range. However, resolving inertial scales in a galactic disk simulation is computationally challenging as it requires high spatial resolution, because turbulence is generated on scales comparable to the disk scale height, $h_{\rm d}\sim 100\pc$, while numerical viscosity affects gas flows in AMR-based codes on scales up to $\sim 10-20$ cells \citep[e.g.,][]{Kritsuk.etal.2011}. Thus, resolving the inertial scales unaffected by numerical viscosity requires minimal cell sizes of $< 5-10\pc$. Therefore, our resolution is not quite within the regime in which the SGS model was calibrated. However, this problem is mitigated by the insensitivity of the SGS turbulence properties in star-forming gas to the parametrization of $\tau_{ij}$, which is the part that depends on the $\Delta$ being within the inertial range. Moreover, as we discuss in Section~\ref{sec:results:GMCs}, we expect SFE to be a weak function of $\Delta$ down to $\sim 1 \pc$ scale. Thus, we believe the use of the SGS turbulence model in simulations with moderate resolution is justified.

Another technical complication relates to the adopted pressure floor that prevents artificial fragmentation of gas \citep[][implemented in our simulations as described in Section~2 of \citealt{Machacek.etal.2001}]{Truelove.etal.1997}. Namely, in regions where cold gas is supported against fragmentation, the artificial pressure set by this floor dominates the flux of the total internal energy ($e+K$) obtained by the Riemann solver. This may result in a non-physical behaviour of internal energy. Therefore, in order to avoid this problem we disregard the internal energy fluxes obtained by the Riemann solver and, instead, at every timestep we advect $e$ and $K$ as passive scalars and then apply all non-conservative source terms. This approach is akin to the dual energy formulation of \citet{Bryan.etal.1995} for protecting internal energy from contamination by numerical errors in highly supersonic flows.

\subsection{Star formation}
\label{sec:method:SFR}

Observational evidence of turbulence in molecular clouds motivated development of analytic models that relate star formation to the properties of self-gravitating MHD turbulence in GMCs \citep{Krumholz.McKee.2005,Padoan.Nordlund.2011,Federrath.Klessen.2012,Hennebelle.Chabrier.2013}. Generally, these models predict variation of star formation efficiency with virial parameter $\avir$ and both sonic and Alfv\'{e}nic Mach numbers. However, they usually rely on strong assumptions about turbulence in GMCs, such as the gas density PDF being static, and the critical density for collapse, that is independent of local flow configuration. Recent direct MHD simulations of turbulent molecular clouds do confirm the strong variation of star formation efficiency with $\avir$ but reveal a surprising insensitivity to other cloud properties \citep{Padoan.etal.2012}. Specifically, \citet{Padoan.etal.2012} find that the star formation efficiency per free-fall time of simulated GMCs can be parametrized by the following simple formula:

\begin{equation}
\label{eq:SFRlaw}
\epsff = \epsw \exp \left( -1.6 \frac{\tff}{\tcr} \right),
\end{equation}
where $\tff = \sqrt{3\pi/32G\rho}$ is a free-fall time that corresponds to initial average density $\rho$ of GMC, $\tcr = \Delta/2\vrms$ is initial turbulent crossing time on the scale of GMC ($\Delta$) and $\epsw$ is a normalization coefficient that takes into account mass loss during formation of stars from protostellar objects modelled by sink particles. Note that ratio $\tff/\tcr$ relates to virial parameter $\avir$, which is usually defined for a uniform sphere with a radius $R$ and mass $M$ as $\avir = 5\sigma^2_{1D}R/GM \approx 1.35 (\tff/\tcr)^2$ \citep{Bertoldi.McKee.1992}. The range of $\tff/\tcr$ probed by \citet{Padoan.etal.2012} covers a wide range of SFE, $\epsff \sim 0.5 - 50\%$, that matches the observed variation, and the above fit holds on the scales of GMCs, few to hundred pc. Therefore, this fit can be directly applied in galaxy formation simulations if the turbulent velocity $\vrms$ on GMC scale is known.

The fit to the numerical results given by Equation~(\ref{eq:SFRlaw}) agrees within a factor of $\sim 2$ with the results obtained by other authors \citep{Clark.etal.2005,Price.Bate.2009,Wang.etal.2010,Krumholz.2012,Federrath.2015}, although such comparison requires care, as SFE is defined and measured differently in different studies.  

Equation~(\ref{eq:SFRlaw}) indicates that SFE is exponentially sensitive to the ratio of self-gravity to turbulent pressure support. Thus, if GMCs have a range of $\avir\propto (\tff/\tcr)^2$ values, the formula implies a wide variation of $\epsff$.

This variation and the relative insensitivity to thermal and Alfv\'{e}nic Mach numbers can be understood as follows. At a fixed sonic Mach number $M$ increasing $\tff/\tcr$ is equivalent to decrease of the average density $\bar{\rho}$ of gas. As the critical density at which gas becomes self-gravitating in physical units is constant, the critical overdensity relative to the average density increases with decreasing $\bar{\rho}$. As a result, the SFE decreases with decreasing fraction of gas mass at overdensities above critical \citep[e.g.,][]{Padoan.Nordlund.2011}. The dependence of SFE on sonic and Alfv\'{e}nic Mach numbers is more complex because their increase results in both widening the density PDF of MHD turbulence and increasing of the critical overdensity \citep{Padoan.Nordlund.2011}. \citet{Padoan.etal.2012} results show that these effects roughly cancel each other and SFE becomes relatively insensitive to the actual values of these Mach numbers, at least at high $M$ explored by these authors. 

In the regime of low-$M$ turbulence the contribution of thermal pressure to the support against gravity cannot be neglected.  To extend the above formula to this regime, we redefine $\tcr$ to take into account thermal pressure support \citep{Chandrasekhar.1951}:

\begin{equation}
\label{eq:tdyn}
\tcr = \frac{\Delta}{2\sqrt{\vrms^2+\cs^2}},
\end{equation}
where $\cs$ is the sound speed.

In our simulations, we estimate $\vrms=\sqrt{2K/\rho}$ from the SGS turbulence energy $K$ and compute SFE in each cell using Equations~(\ref{eq:SFRlaw}) and (\ref{eq:tdyn}) with the cell size for the value of $\Delta$. We adopt $\epsw = 0.9$ consistent with the results of \citet{Federrath.etal.2014} who showed that the mass loss due to outflows does not exceed $10\%$ on the scale where \citet{Padoan.etal.2012} form sink particles.

We stress that all parametrizations and parameter values in our star formation model are not tuned but taken from the results of simulations of star formation in GMCs.

\subsection{Stellar feedback}
\label{sec:method:feedback}

Formation of stars is accompanied by stellar feedback: i.e. injection of mass, momentum, and energy into surrounding gas. Recent studies \citep[e.g,][]{Stinson.etal.2013,Hopkins.etal.2014,Agertz.Kravtsov.2015} have demonstrated the importance of stellar feedback in shaping the key galaxy properties, such as stellar mass, size, morphology, and star formation history. The challenge of modelling feedback in galaxy formation simulations is due to the fact that the relevant processes are not fully understood and that any implementation needs to work at the resolution limit of simulations. 

In our simulation, we calculate a total number of SNe II associated with a given young stellar particle assuming the Miller-Scalo IMF \citep{Miller.Scalo.1979}. We assume that SNe explode at a uniform rate between $3$ and $40\Myr$ after the stellar particle is created, with each supernova releasing $10^{51}$ ergs. We use the results of SN remnants simulations by \citet{Martizzi.etal.2015} to estimate the fraction of this energy that gets converted into the momentum of expanding shell (their Equation~(10)) and the fraction that is left in form of thermal energy at the snowplow stage (their Equation~(9)), when the size of the SN remnant reaches the resolution scale $\Delta$. 

Furthermore, given that in a star-forming region the remnants of multiple SNe can collide, their coherent expansion may transform into turbulent gas motions with partial momentum cancellation. As GMCs typically have distinct filamentary structure, we expect that $\sim 1/3$ of the total momentum is transformed into turbulent motions. Therefore, $65\%$ of the total radial momentum given by Equation~(10) of \citet{Martizzi.etal.2015} we assign as a radial momentum to 26 neighbors of the stellar particle-host cell, while $30\%$ is converted into the SGS turbulent energy via the term $S_{\rm SN}$ in our Equation~(\ref{eq:K}). Finally, we add $5\%$ of the initial SN energy to the thermal energy of the host cell.

\subsection{Galaxy model and numerical resolution}
\label{sec:method:ICs}

In this work we follow the evolution of a Milky Way-like disk galaxy consisting of gaseous and stellar disks with a bulge embedded in a spherical dark matter halo. We use initial conditions for the stellar component and halo similar to those used by \citet{Agertz.etal.2013}, which are also similar to the initial conditions for the disk simulations in the AGORA comparison project \citep{Kim.etal.2014}. 

Specifically, the dark matter halo profile has the NFW form with concentration of $c=10$ and circular velocity at the radius enclosing the density contrast of 200 relative to the critical density of the universe of $v_{200}=150 \kms$ \citep{NFW.1997}. For the baryonic disk we use an analytical exponential density distribution with a scale radius $r_{\rm d}\approx3.4\kpc$ and scale height $h_{\rm d}=0.1r_{\rm d}$. The spherical stellar bulge is described by the Hernquist profile \citep{Hernquist.1990} with a scale radius $r_{\rm b}=0.1r_{\rm d}$ and bulge to total disk mass ratio $M_{\rm b}/M_{\rm d} = 0.1$. 

The total mass of the baryonic disk is similar to that of the Milky Way: $M_{\rm d}\approx 4.3\times10^{10} \Msun$ with $20\%$ of mass in the gaseous disk and the rest in the stellar component. This gas mass fraction is slightly higher than the observed current $\sim13\%$ in the Milky Way \citep{Nakanishi.Sofue.2015}, but is perhaps appropriate at a somewhat earlier stage of its evolution, about $1 \Gyr$ ago.  

During the disk evolution we adaptively refine cells according to a gas mass criterion. Specifically, we split a cell when its total gas mass exceeds $8.3\times10^3\Msun$. By doing so we reach a maximal spatial resolution of $\Delta = 40\pc$. This scale is similar to the typical resolution of modern cosmological zoom-in simulations. In addition, this scale corresponds to the size of the largest observed GMCs. Given that the star formation prescription was calibrated on the GMC scales, we decide to stop at this relatively coarse resolution and study the properties of this recipe in the regime applicable to cosmological runs. In addition to this fiducial run, we also rerun our simulation at higher resolutions, $\Delta = 20$ and $10\pc$, and briefly describe these results in Sections~\ref{sec:results:GMCs} and \ref{sec:results:KS}. We will present more detailed comparison in a follow-up paper.

We set the initial ambient halo gas density and temperature to $10^{-6}\cc$ and $10^4\K$ respectively. As a result, the exponential gaseous disk with a constant scale height initially is slightly off thermal equilibrium. To make the transition to equilibrium state more gradual we initialize the subgrid turbulent velocities in the gas uniformly at the value of $\vrms = 1 \kms$ to provide an additional pressure support. We checked that the resulting distribution of turbulent velocities in the evolved disk does not depend on the initial value of $\vrms$.

To further minimize transient effects related to initial disk self-adjustment, we begin the evolution of the disk with cooling, star formation and the SGS turbulence model off. We allow the disk to settle into an equilibrium and turn on dynamical production and dissipation of the SGS turbulence after $50\Myr$. Cooling is then turned on gradually between $100\Myr$ and $200\Myr$. After that, we evolve the disk for another $100\Myr$ allowing all remaining transient effects to dissipate. At this point, at $t=300\Myr$ from the beginning of the simulation, the disk is in a steady thermal and dynamical equilibrium and maintains its structure, velocity and density profiles over further evolution if the star formation remains switched off. Thus, we turn on the star formation at $t=300\Myr$.

For our analysis we choose a single snapshot of the disk evolution at $t \approx 600\Myr$, i.e. $300\Myr$ after the star formation is turned on. This choice is a compromise between allowing the disk to evolve for an appreciable amount of time in order to settle into a steady state, on the one hand, and choosing a time significantly shorter than the global gas depletion time ($\tdep \sim 1\Gyr$), on the other hand. Given that our simulation does not include explicit modelling of the cosmological mass accretion, the latter condition ensures that we examine the disk before it significantly exhausts its gas supply, which could bias our comparisons with observations.

\section{Results}
\label{sec:results}

\begin{figure*}
\centering
\includegraphics[width=\textwidth]{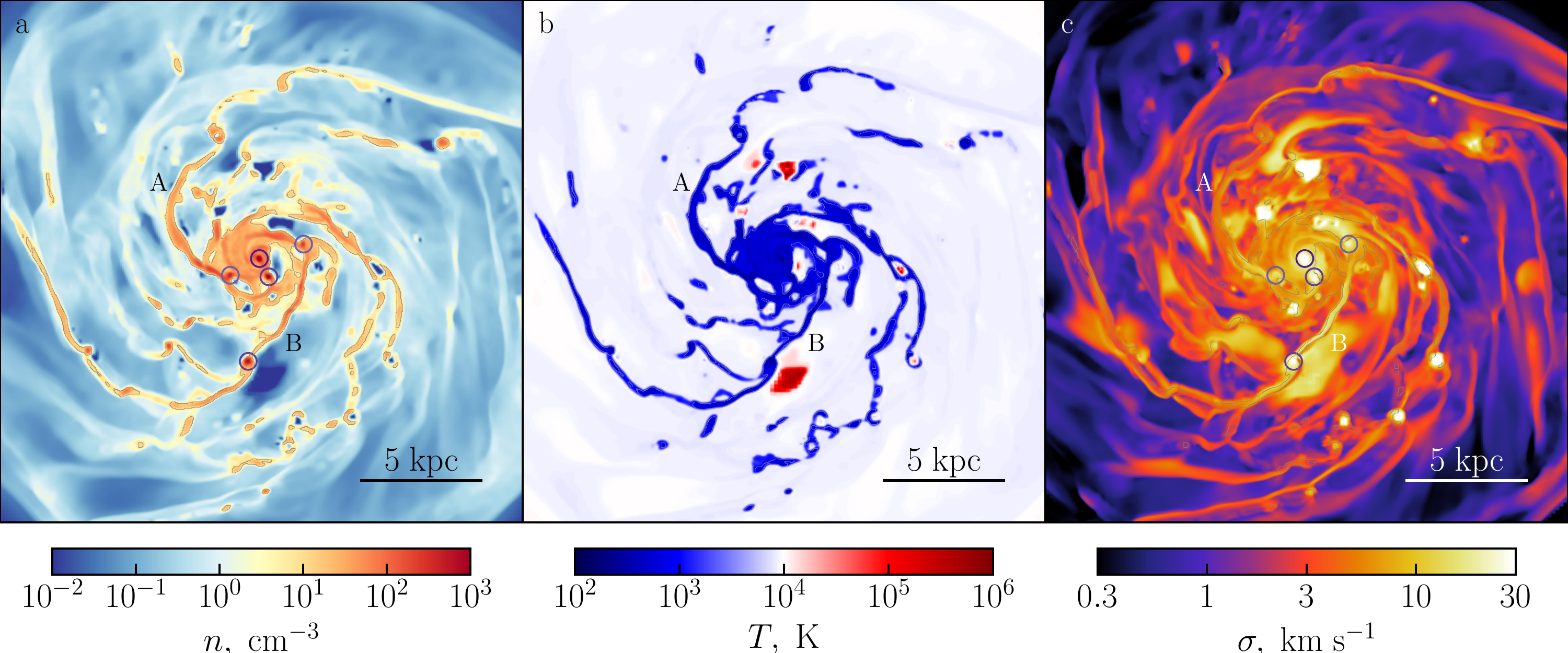} 
\vspace*{0.5ex} \\
\includegraphics[width=\textwidth]{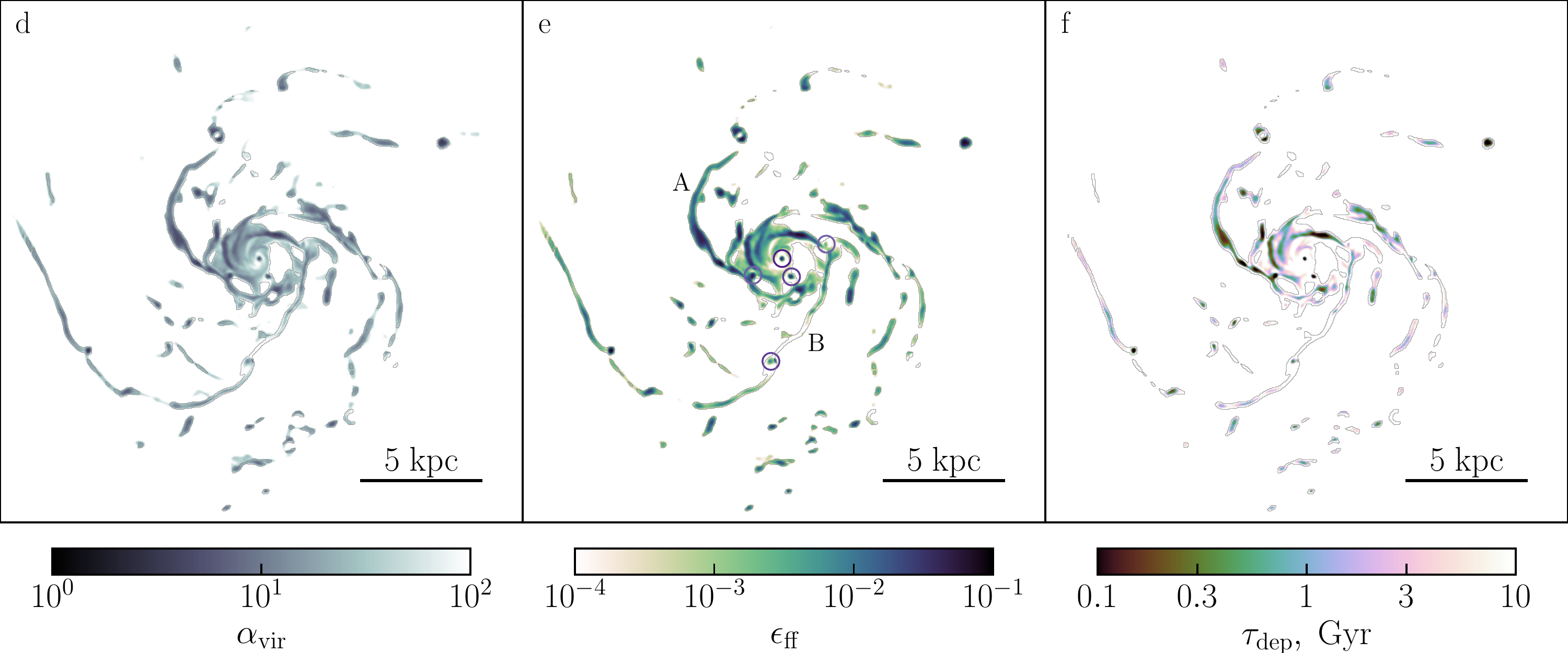}
\caption{\label{fig:1:disk} Gaseous disk after $600\Myr$ of evolution. The \textit{top row} shows from left to right slices of density (a), temperature (b) and subgrid rms turbulent velocity (c) in the disk plane. The temperature and rms velocities are derived from the thermal and SGS turbulent energies respectively. The \textit{bottom row} shows derived quantities related to the star formation prescription. \textit{Left panel} (d) shows the virial parameter calculated under the assumption that clouds are uniform spheres of mass $M$ and radius $R$: $\avir = 5\sigma^2_{1D}R/GM \approx 1.35 (\tff/\tcr)^2$ \citep{Bertoldi.McKee.1992}. Only the gas mass was taken into account in estimation of $\avir$. $\tff$ and $\tcr$ are derived from the density and turbulent velocity shown in panels a and c. Equation~(\ref{eq:SFRlaw}) translates the derived $\tff/\tcr$ directly into SFE shown in the \textit{middle panel} (e). Thin grey lines in all six panels indicate an iso-density contour that corresponds to $n=10\cc$ and, therefore, encompasses cold dense gas. Thus, the predicted SFE exhibits strong spatial variation even in the cold gas. \textit{Right panel} (f) shows the distribution of the gas depletion time defined as $\tdep \equiv \rho/\rhosfr$. Purple circles in panels a, c and e indicate dense gaseous clumps.}
\end{figure*}

The star formation efficiency in our model is predicted based on the self-consistently evolved local density, temperature and subgrid turbulent velocity. Thus, in this section we discuss the distribution of these quantities in our simulated disk with a special emphasis on the turbulent velocities. We also compare to observations the star formation efficiencies and rates predicted by our turbulence-based star formation prescription.  

As can be seen in panels a and b of Figure~\ref{fig:1:disk}, by $t\approx 600$ Myr most of the disk volume is filled by the diffuse ($n\sim0.1-2\cc$) warm ($T\sim10^4\K$) gas, while denser gas resides in spiral structures. The spiral arms travel around the disk compressing diffuse gas for certain periods of time. Their motion relative to the diffuse gas is generally supersonic and these spiral waves are thus accompanied by shocks. As gas is being compressed, at $n\sim5\cc$ cooling by thermal excitation of CII and OI fine structure lines becomes efficient and the gas rapidly cools down. This substantial cooling in spiral arms means that the pre-arm shocks are radiative and, therefore, they produce density jumps of orders of magnitude at the spiral arm interfaces. 

The highest densities are reached in self-gravitating gaseous clumps. These clumps develop within the spiral arms due to local gravitational instabilities \citep[e.g.,][]{Agertz.etal.2009b,Dekel.etal.2009,Bournaud.etal.2010}. Several examples of such clumps are circled in Figure~\ref{fig:1:disk}a. In contrast to spiral arms, these high-density clumps are persistent physical objects rather than waves and so they may survive for a significant period of time. Examination of disk evolution shows that some of the clumps last up to a couple disk revolutions, until they are disrupted by feedback or merged with the gas concentration in the disk center, as also found in a number of other recent studies \citep[e.g.,][]{Genel.etal.2012}. The long-lasting clumps may themselves drive the formation of the disk spiral structure \citep[e.g.,][]{DOnghia.etal.2013}.

Hot rarefied bubbles of gas are another kind of prominent features seen in Figures~\ref{fig:1:disk}a and b. Some of these bubbles are inflated by exploding SNe in the regions with active star formation. Local injections of SNe energy and momentum affect the distribution of dense cold gas, as they disrupt gaseous clumps and tear spiral arms apart. Sometimes, as in the case of the large hot spot near the marker ``B'', the hot gas instead is being pushed into the disk plane from the hot halo ($T\approx T_{\rm vir}\sim 10^6\K$) in regions where the disk is thinned and its gas pressure is low.

\subsection{Properties of the ISM turbulence}
\label{sec:results:turbulence}

\begin{figure}
\centering
\includegraphics[width=\columnwidth]{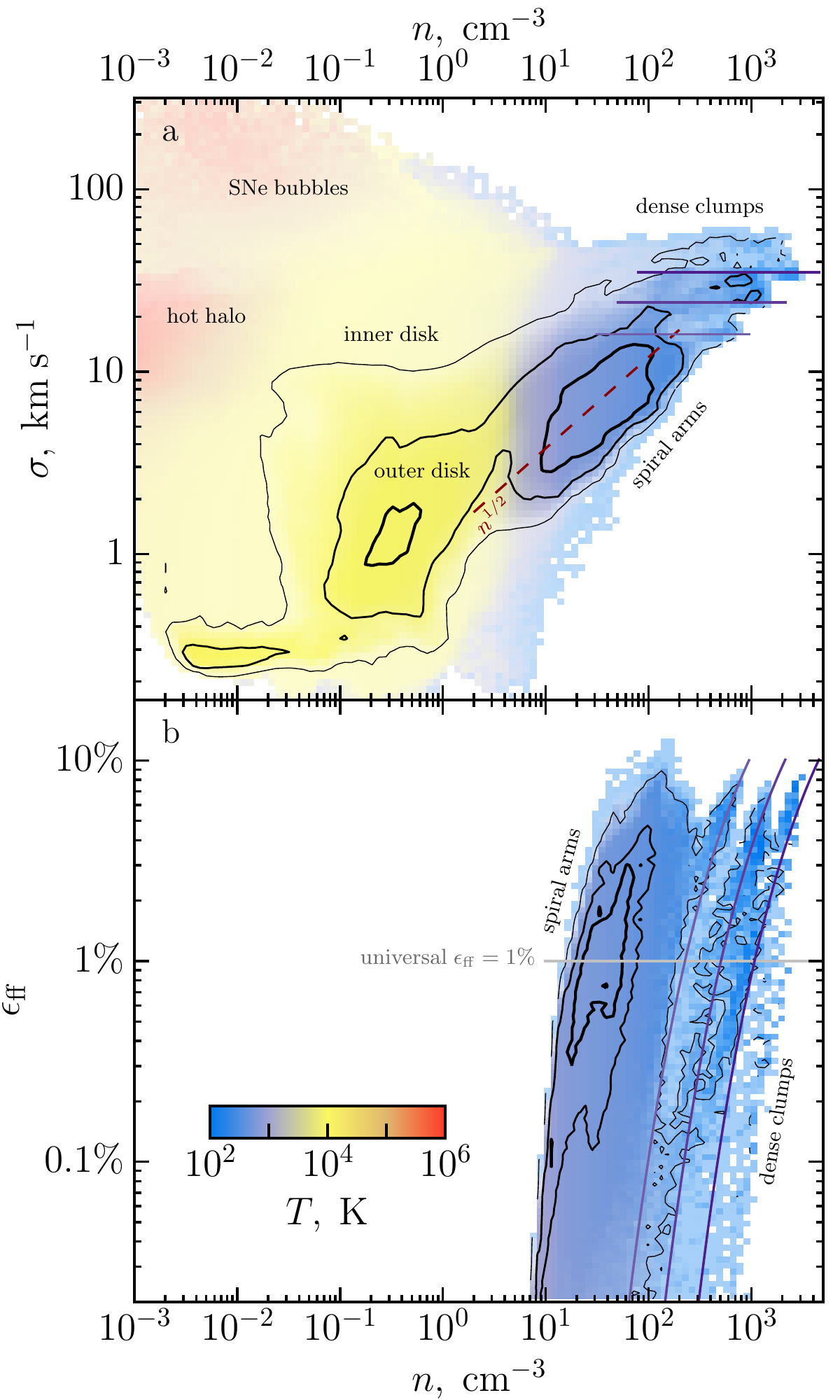}
\caption{The distribution of SGS turbulent velocities $\vrms$ (\textit{top panel}) and the resulting star formation efficiencies per free-fall time $\epsff$ (\textit{bottom panel}) at different densities. The distributions take into account all cells within cylindrical volume with $R<20\kpc$ ($\sim 6$ initial scale radii) and $|z|<1\kpc$ (total height is $\sim 6$ initial scale heights) centered at the disk center. To increase statistics we average PDFs over 23 snapshots at $600 \pm 10\Myr$. Colors show the mass-weighted average temperature in bin and its intensity indicates the total mass in bin. Black contours enclose 25\%, 68\%, 95\% (\textit{top}) and 5\%, 15\%, 30\% (\textit{bottom}) of the current total gas mass. The turbulent velocities in cold spiral arms result from compression of diffuse gas and scale with density roughly as $\vrms \propto n^{1/2}$ (dashed red line given by Equation~(\ref{eq:sigma})). The linear structures at the upper right end of the $n-\vrms$ distribution correspond to dense gaseous clouds with $\vrms \approx \rm const$. \textit{Bottom panel} shows the distribution of star formation efficiency obtained from the local gas density, temperature and SGS turbulent velocities. The adopted model naturally introduces an exponential cutoff at densities $n \sim 10 \cc$. If compared to universal $\epsff$ prescriptions turbulent model predicts broad variation by orders of magnitude, even though the average SFE is still $\sim 1\%$.}
\label{fig:2:phases} 
\end{figure}

Given the importance of turbulence for our adopted star formation prescription, we begin with the discussion of the small-scale subgrid turbulence in the disk. Figure~\ref{fig:1:disk}c shows that the subgrid model predicts the rms turbulent velocities, $\vrms$, at the level of few to ten$\kms$ on the scale of our smallest grid cells ($40\pc$). This result agrees with the observed velocity dispersion in GMCs \citep{Gammie.etal.1991,Bolatto.etal.2008}, Milky Way dynamics \citep{Kalberla.Dedes.2008} and extragalactic HI data \citep{Petric.Rupen.2007,Tamburro.etal.2009}, as well as with high resolution disk simulations \citep{Agertz.etal.2009}. The turbulent velocities in our simulation increase towards the disk center where gravitational instabilities and frequent SNe maintain higher $\vrms$ \citep{Agertz.etal.2009}. High $\vrms$ in bright spots that correspond to hot gas ($10^5-10^6\K$ in Figure~\ref{fig:1:disk}b) are driven by expanding supernova bubbles. Dense cold spiral arms are typically more turbulent than the surrounding gas and therefore they are well traceable in the $\vrms$ map, especially at $r>5\kpc$. Enhanced turbulent velocities in the spiral arms result from the compression of inter-arm turbulence. Similarly, collapse of gas into dense gaseous clumps also results in high turbulent velocities (circled in Figures~\ref{fig:1:disk}a and c). 

Quantitative conclusions about turbulence in different ISM phases can be drawn from the distribution of $\vrms$ as a function of local gas density shown in Figure~\ref{fig:2:phases}a. In this plot several distinct phases are highlighted using color: the warm diffuse gas at $T\sim 10^4 \K$ (\textit{yellow}), the cold dense gas in spiral arms and dense clumps (\textit{blue}), and the hot tenuous gas at $T>10^5\K$ in the SNe bubbles and hot gaseous halo surrounding the disk (\textit{red}). The contours enclosing different mass fractions show that most of the gas mass is in the warm and cold phases.

The warm gas phase corresponds to the diffuse gas between spiral arms and around the disk plane (white color in Figure~\ref{fig:1:disk}b). In this phase the turbulence is in an approximate equilibrium between production due to instabilities \citep[e.g.,][]{Bournaud.etal.2010}, sourcing by SNe and viscous dissipation into heat. Most of the gas mass in this phase resides on the disk outskirts ($r>5\kpc$, labelled as the ``outer disk'') and the typical {\it subgrid} turbulent velocities are $\sim 1-2\kms$ with a significant scatter of $\sim 0.5\dex$. Note that the actual velocity dispersion of the disk would include resolved gas motions that are considerably larger. As mentioned above, $\vrms$ in the diffuse gas increases towards the disk center and may reach few tens to hundred$\kms$. Some of the warm gas with the largest velocity dispersions resides in expanding SNe bubbles, which drive turbulent velocities to the highest values found in our simulations, few hundred$\kms$.

The cold gas phase is, of course, the most interesting for star formation. The figure shows that such gas has typical subgrid turbulent velocities of $\vrms\sim3-10\kms$. This result agrees with the observed three-dimensional turbulent velocities in GMCs \citep[\mbox{$\sim 8 \kms$} on scales of $40\pc$ in][]{Bolatto.etal.2008}. We find that the actual value of $\vrms$ correlates with the local compression rate ($-\nabla_i u_i$). This result indicates that the main source of turbulent energy in this regime is heating by compression of the diffuse gas \citep{Robertson.Goldreich.2012}. Specifically, as a parcel of relaxed gas at $T\sim10^4\K$ enters a spiral arm both thermal and turbulent energies increase, as pressures associated with thermal and random motions do negative work during compression. However, at typical spiral arm densities the excess of thermal energy is quickly radiated away and the gas cools down. In contrast to thermal energy, turbulent energy dissipates on the local crossing time scale, $t\sim \Delta/\vrms$ \citep[e.g.,][]{MacLow.etal.1998}, which may be longer than the time spent by the gas parcel inside the spiral arm. The turbulent dissipation time scale and the time spent in the spiral arm can be estimated as:

\begin{align}
\label{eq:tdec}
\tdec &\approx 4\Myr\left(\frac{\Delta}{40\pc}\right)\left(\frac{\vrms}{10\kms}\right)^{-1}, \\
\label{eq:tarm}
\tarm &\approx 3\Myr\left(\frac{\warm}{300\pc}\right)\left(\frac{\varm}{100\kms}\right)^{-1},
\end{align}
where $\warm$ is a typical spiral arm width and $\varm$ is its typical velocity relative to the ambient gas. Here we neglect the fact that gas may enter into spiral arms at different angles and then travel along the spiral arm. We approximate the spiral arm passing time simply as $\warm/\varm$ with typical values found in our simulation. Actual $\tarm$ may vary around the estimation from Equation~(\ref{eq:tarm}) depending on the local gas dynamics.

In the outer disk, where the pre-shocked gas has low turbulent velocities ($\vrms \sim 1\kms$), turbulence decays slowly ($\tdec > \tarm$) and $\vrms$ increases in spiral arms. The bimodal distribution of mass in Figure~\ref{fig:2:phases}a indicates that the compression is fast and, therefore, the gas is either relaxed or resides in a spiral arm. In the inner disk where $\vrms$ is high ($\sim 10\kms$) turbulence may decay during compression ($\tdec \sim \tarm$) and the increase of $\vrms$ with density is shallower. Turbulence decays even more efficiently in the dense gaseous clumps where $\vrms$ reaches $\mathrm{few\ tens}\kms$. Their typical lifetime ($>100\Myr$) is considerably longer than the turbulence decay time scale (few$\Myr$ from Equation~(\ref{eq:tdec})). As a result, $\vrms$ reaches an equilibrium value, that weakly depends on density (purple lines in Figure~\ref{fig:2:phases}). 

Star formation in our model proceeds in the cold dense gas. The pressure support in this gas is dominated by small-scale turbulent motions, as the sound speed in this phase is $\cs \sim 1 \kms \sqrt{T/100\K}$ and the turbulent velocities ($\vrms > 3\kms$) are supersonic. Therefore, the gas in this regime forms stars with the efficiency that depends on $\vrms$ (Equations~(\ref{eq:SFRlaw}) and (\ref{eq:tdyn})). In our simulation we find that the average $\vrms$ in cold gas depends on density as (dashed red line in Figure~\ref{fig:2:phases}a): 

\begin{equation}
\label{eq:sigma}
\vrms = 12\kms \left( \frac{n}{100\cc} \right)^{1/2}.
\end{equation}

This scaling with density is consistent with compression in a ``non-radiative'' (with respect to turbulent energy) shock. In this case the turbulent ``temperature'' after the shock goes as the Mach number squared: $\vrms^2 \propto M^2$. As noted before, the associated orders of magnitude density jumps indicate that the shocks preceding spiral arms are almost isothermal (with respect to thermal energy). In this case the density jump also goes as the Mach number squared: $\rho \propto M^2$ \citep[e.g.,][]{Zeldovich.Raizer}. These scaling relations for $\vrms$ and $\rho$ then imply $\vrms \propto \rho^{1/2}$. 

The obtained scaling is opposite to what follows from the empirical Larson's relations, $\vrms \propto \rho^{-0.5}$ \citep{Larson.1981}, that are also consistent with the turbulent origin \citep{Kritsuk.etal.2013}. This difference originates from the mismatch of scales on which we consider turbulence with the observed GMCs sizes. Most of the observed GMCs have sizes of $\sim 1-10\pc$ and therefore they reside in the inertial interval of turbulent cascade driven on the scale of disk height, $\sim 200-300\pc$. At the resolution of our simulation the inertial range is not yet resolved (see Section~\ref{sec:method:SGST}), and thus, we cannot expect correct predictions of the Larson's relations. The turbulent cascade that is expected to develop on the scales unresolved in our simulations would result in an inverted relation between $\vrms$ and $\rho$ consistent with the Larson's relation \citep{Kritsuk.etal.2013}.

Figure~\ref{fig:2:phases}a shows significant scatter around the average behavior expressed by Equation~(\ref{eq:sigma}) ($\sim 0.3$ dex). We find that this scatter is mostly due to the variation of local compression rate ($-\nabla_i u_i$). This offers hope that $\sigma$ may be approximated in simulations without explicit SGS turbulence modelling using dependencies of $\vrms$ on density and $\nabla_i u_i$, that can be calibrated using simulations with such modelling. We will explore the relation of the local compression rate and the subgrid turbulent velocity in a future study.

In closing we note, that the above discussion shows that the key mechanism of turbulence production in star-forming regions of our disk is the compression of warm, transonic gas by spiral waves. This justifies the usage of the linear closure for the turbulent stress tensor $\tau_{ij}$, as discussed in Section~\ref{sec:method:SGST}. In particular, production of SGS turbulence from resolved motions is mostly important in diffuse gas with $T\sim10^4\K$. This temperature corresponds to the sound speed of $\cs\sim 10\kms$ and, therefore, the typical turbulent velocities in this gas ($\mathrm{few}\kms$) are sub- or transonic, for which the linear closure for the stress tensor is more appropriate \citep{Schmidt.Federrath.2011}.

\subsection{Star formation efficiency}
\label{sec:results:SFR}

In our simulation we derive the star formation efficiency $\epsff$ in each cell using Equations~(\ref{eq:SFRlaw}) and (\ref{eq:tdyn}). These equations parametrize local SFE via virial parameter:
\begin{equation}
\avir \propto \left(\frac{\tff}{\tcr}\right)^2 \propto \frac{\vrms^2+\cs^2(T)}{\rho},
\end{equation}
where $\rho$, $T$ and $\vrms$ are the gas density, temperature, and SGS turbulent velocity dispersion self-consistently evolved by the code. As can be seen in Figure~\ref{fig:1:disk}d the virial parameter of modelled cells on scale of $40\pc$ is rather high and exhibits significant variation. 

Given the exponential dependence of $\epsff$ on the virial parameter, the $\avir$ variation translates into even wider spatial variation of $\epsff$. This variation in the dense gas can be seen in panels e and f of Figure~\ref{fig:1:disk}, that show maps of $\epsff$ and the gas depletion time to star formation relative to the gas denser than $n=10\ \rm cm^{-3}$, shown by the thin gray contours.

The depletion time of molecular gas, and thus possibly $\epsff$, is indeed observed to vary along the spiral arms of M51 \citep{Meidt.etal.2013}. As an extreme example of this spatial variation compare the spiral arms denoted as A and B in the panel e of Figure~\ref{fig:1:disk}. Panels a and b of the figure show that the gas in these arms has similar density and temperature and, therefore, universal SFE model would predict the same efficiency and similar depletion time in both arms. However, in our simulation the spiral arm A forms stars much more efficiently than the arm B due to lower turbulent velocity predicted by the SGS model. 

As discussed in Section~\ref{sec:results:turbulence} the difference in $\vrms$ originates from the variation in the local compression rates. The compression rate, in its turn, may vary due to several reasons. First, gas may experience different compression in spiral arms depending on the large-scale dynamics and development of local disk instabilities. Second, turbulence may be suppressed (enhanced) in spiral arms due to local expansion (contraction) of gas along the arm. Third, spiral arms may be affected by hot gas from either SNe bubbles or the halo gas penetrating into the disk. In particular, as can be seen in Figure~\ref{fig:1:disk}b, the spiral arm B is adjacent to a bubble of hot gas in the downstream direction. The thermal pressure of this hot gas may contribute to compression. 

In all three scenarios higher turbulent velocities result from stronger compression. More quantitative information about the SFE variation with density can be drawn from the phase diagram shown in Figure~\ref{fig:2:phases}b. The most noticeable features of this diagram are the sharp cutoff at $n \sim 10\cc$ and the orders of magnitude variation of $\epsff$.

The exponential cutoff at $n \sim 10\cc$ in our model arises naturally from the thermal support in warm diffuse gas. In particular, turbulent velocities in this phase are mostly subsonic and gas is supported against gravity mainly by its thermal pressure. This thermal support is codified in the definition of the crossing time given by Equation~(\ref{eq:tdyn}), which results in the exponential suppression of $\epsff$ in the diffuse gas, as the sound crossing time becomes considerably shorter than the free-fall time: $\tff/\tcr \sim 26 \sqrt{T_4/n_0}$, where $T_4 = T/10^4 \K$ and $n_0 = n/1 \cc$. 

In the cold phase, on the other hand, the turbulence is supersonic and its pressure provides the main support against gravity. The typical turbulent crossing time in this regime is of the order of the free-fall time: $\tff/\tcr \sim 2.6 \vrms_{1}/\sqrt{n_{2}}$, where $\vrms_1 = \vrms/10 \kms$ and $n_{2} = n/100 \cc$. This ratio corresponds to  $\epsff \sim 1\%$, typical for star-forming regions. As a result, in the turbulent model only cold dense gas forms stars at a reasonably large efficiency. The transition from the negligible values of $\epsff$ in the warm diffuse gas to $\epsff\sim 1\%$ in the cold dense gas is sharp due to the abrupt drop in temperature. This sharp transition is responsible for the effective density threshold for efficient star formation at $n \sim 10 \cc$.

The most efficient star formation in our simulation occurs in the gas of density $n\sim10-100\cc$ in spiral arms and dense clumps. The average trend of $\vrms$ with density ($\vrms \propto n^{1/2}$ from Equation~(\ref{eq:sigma})) substituted into Equation~(\ref{eq:SFRlaw}) results in constant $\epsff \sim 0.6\%$ independent of density, as $\avir \propto \vrms^2 n$. Therefore, the entire variation of $\epsff$ around the average value originates from the scatter of modelled turbulent velocities around the average trend. As we mentioned before, this scatter is related to the variation of local compression rate.

Although the mass-weighted average $\epsff$ in our disk is quite similar to the universal value $\epsff \sim 1\%$ at $n > 10\cc$ usually inferred from observations, the large spatial and temporal variation of the SFE predicted by our model may have important effects on galaxy evolution. For example, localization of efficient star formation in high-density regions may have a drastic effect on the ability of galaxy to drive large-scale winds and affect the final morphology of the galaxy \citep[e.g.,][]{Governato.etal.2010}.

\subsection{Comparison with observed GMCs}
\label{sec:results:GMCs}

\begin{figure}
\centering
\includegraphics[width=\columnwidth]{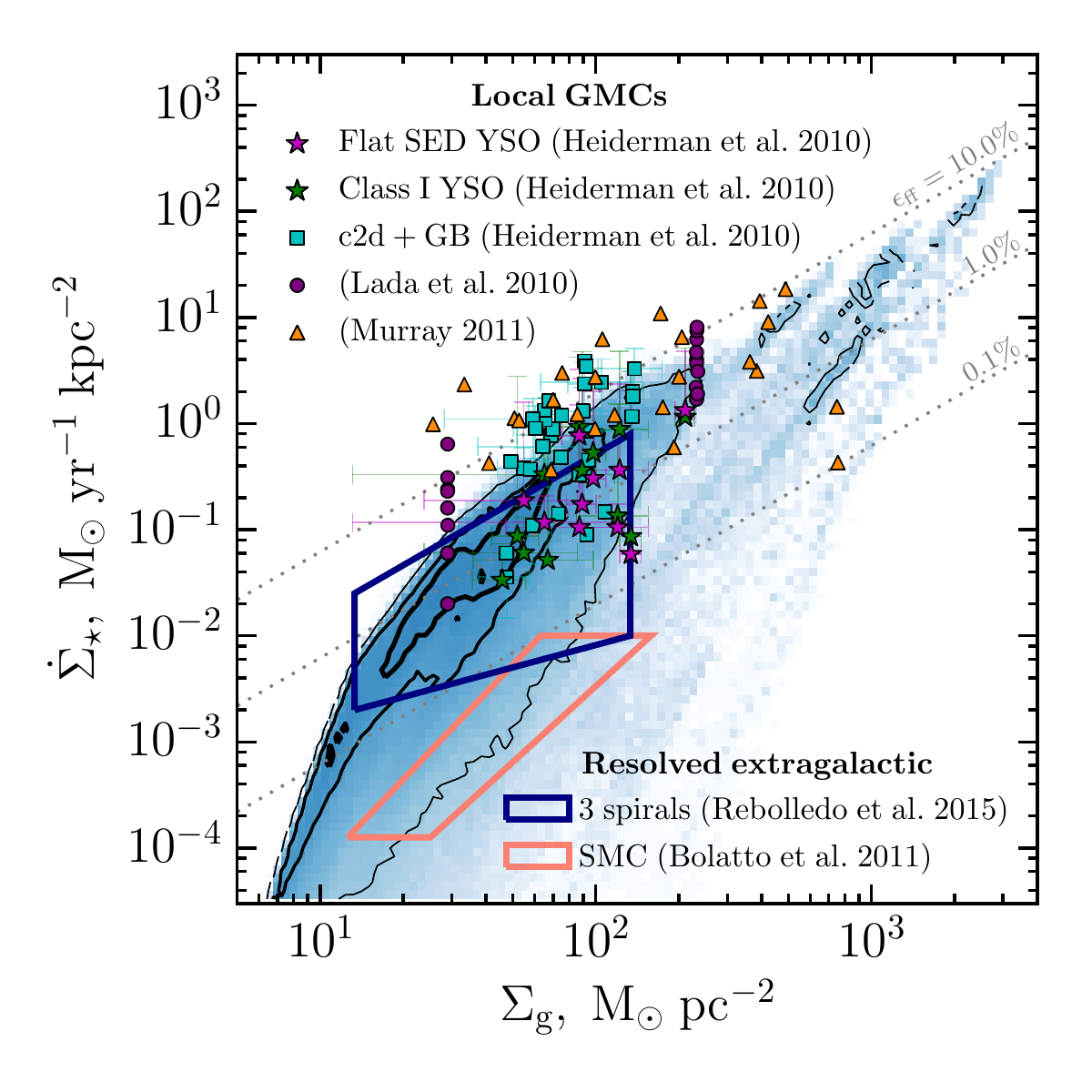}
\caption{Star formation rates obtained in our simulation are broadly consistent with observational data. Blue color indicates mass weighted distribution of cells in $\Sgas - \Ssfr$ plane. Black contours indicate 5\%, 15\% and 30\% of the current total gaseous disk mass. Grey dotted lines correspond to constant values of $\epsff$ in the simulated star-forming regions. The overplotted data points show different samples of GMCs in the Milky Way: \citet[stars and squares]{Heiderman.etal.2010}, \citet[circles]{Lada.etal.2010} and \citet[triangles]{Murray.2011}. In this plot we show only GMCs with sizes in the range $\sim 5-100\pc$ that roughly correspond to our cell size, $\Delta = 40\pc$. Two polygons show resolved star formation rates in nearby galaxies. The blue one summarizes results of \citet{Rebolledo.etal.2015} for three nearby spirals: NGC 6946, NGC 628 and M101, while the pink one indicates star formation in the Small Magellanic Cloud \citep{Bolatto.etal.2011}. In the \citet{Rebolledo.etal.2015} sample we correct gas surface densities for helium assuming 25\% mass fraction.}
\label{fig:3:GMCs} 
\end{figure}

In order to check the viability of our star formation model, we compare its predictions to the SFRs in observed GMCs. Specifically, in Figure~\ref{fig:3:GMCs} we compare the local gas surface density ($\Sgas = \rho \Delta$) and the surface density of SFR ($\Ssfr = \rhosfr \Delta = \epsff \rho \Delta/\tff$) in individual cells ($\Delta = 40\pc$) to the corresponding quantities measured in GMCs. For a fair comparison, we select the observed clouds with sizes in the range $\sim 5-100\pc$, that straddle the cell size in our simulation. 

Distribution of the SFR in our disk has a sharp upper boundary with a wide tail towards lower rates. The observed local star-forming regions from \citet{Heiderman.etal.2010,Lada.etal.2010,Murray.2011} agree remarkably well with the upper envelope of our predicted distribution, whereas the extragalactic data agrees with the main mode of SF in our simulation. This may be because studies of the local GMCs focus on the regions with the most efficient star formation. Blind surveys of star formation in the Milky Way GMCs, on the other hand, start revealing abundant gas with $\epsff\ll 1\%$ (N. Murray, private communication). 

In contrast to the local surveys, GMCs in other galaxies are sampled more uniformly and do indicate prevalent dense gas with low star formation efficiency. For instance, the blue polygon in Figure~\ref{fig:3:GMCs} summarizes the \citet{Rebolledo.etal.2015} results for three nearby spiral galaxies: NGC 6946, NGC 628 and M101. Their inferred SFRs do agree with the typical SFRs of the dense gas in our simulation. The observed SFR distribution in the Small Magellanic Cloud \citep[SMC; pink polygon in Figure~\ref{fig:3:GMCs};][]{Bolatto.etal.2011} also reveals that most of its dense molecular gas forms stars rather inefficiently. However, the SMC is a dwarf galaxy with substantially different dynamics and significantly lower metallicity than the Milky Way, and therefore, its global SFR is considerably lower than that of our simulated galaxy. This discrepancy is yet another manifestation of the SFE variation and its dependence on galaxy properties.

As we discussed in Section~\ref{sec:results:turbulence}, there is a mismatch of our resolution scale with the sizes of the observed GMCs. This may be a cause of concern for the comparison presented above. However, our convergence study indicates that the agreement with observations persists at higher resolutions. At smaller $\Delta$, denser and denser structures are being resolved, while turbulent velocities in these structures keep following average trend given by Equation~\ref{eq:sigma}, $\vrms \propto n^{1/2}$, resulting in similar $\epsff$ distributions. As a consequence, in higher resolution runs the PDF shown in Figure~\ref{fig:3:GMCs} shifts to higher surface densities along $\epsff = {\rm const}$ lines and remains in good agreement with both local and extragalactic data. However, at some point, when the turbulent inertial range becomes fairly resolved, we expect $\vrms - n$ trend to reverse to the observed Larson's scaling, $\vrms \propto n^{-0.5}$. This will cause further motion of the PDF towards the observed data points perpendicular to the $\epsff = {\rm const}$ lines. In other words, the obtained scaling of $\vrms$ with density makes star formation efficiency a weak function of scale down to the scale of individual GMCs in the inertial regime. This weak dependence justifies the comparison of SFRs on the scale of individual cells in our simulation to the observed SFRs on GMC scale.

\subsection{Comparison with observed Kennicutt-Schmidt relations}
\label{sec:results:KS}

\begin{figure}
\centering
\includegraphics[width=\columnwidth]{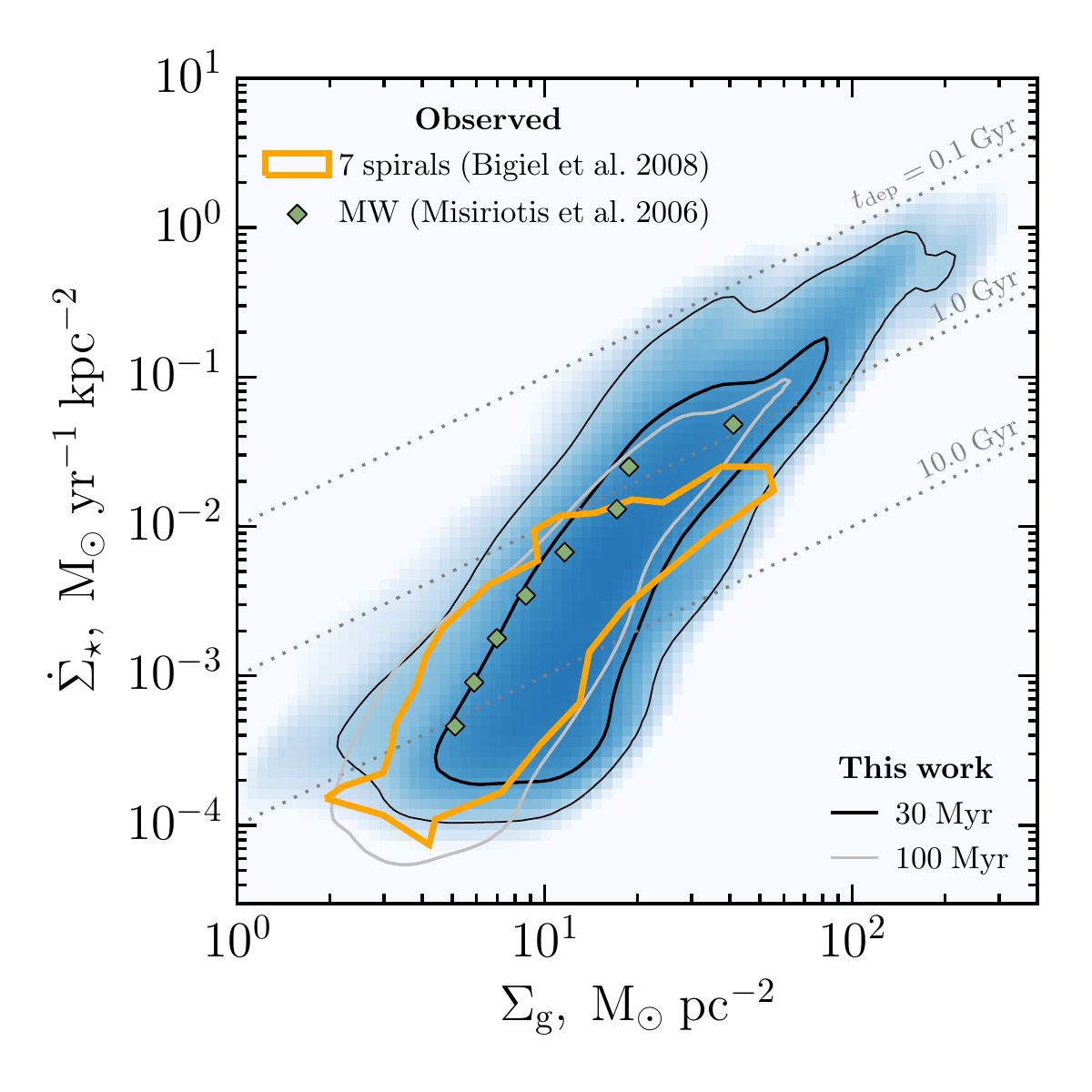}
\caption{Our predicted averaged SFRs agree with observed Kennicutt-Schmidt relations. Blue color and black contours (68\% and 95\% c.l.) indicate smoothed distribution of vertically integrated SFR and gas density in \mbox{$1\times1\kpc$} patches. To calculate SFR in each patch we divide its total stellar mass formed over the last $t_{\rm SFR} = 30\Myr$ of the disk evolution by the time scale $t_{\rm SFR}$. For comparison we also plot a single grey contour (68\% c.l.) obtained for $t_{\rm SFR} = 100\Myr$. We compare our results to the observed SFRs on $750\pc$ scale in 7 nearby spirals \citep{Bigiel.etal.2008} and measurements of the SFR on comparable scale in the Milky Way \citep{Misiriotis.etal.2006}. In the observational data we correct the gas surface densities for helium assuming 25\% mass fraction. Grey dotted lines correspond to constant values of $\tdep$.}
\label{fig:4:KS} 
\end{figure}

The distribution of local SFR in galaxies averaged over certain spatial and temporal scales results in the observed power-law relation between SFR and gas surface density, the Kennicutt-Schmidt (KS) relation \citep[][]{Schmidt.1959,Schmidt.1963,Kennicutt.1998}. When averaged over entire galaxies the KS relation indicates the global gas depletion time of molecular gas of $\tdep\equiv\rho/\rhosfr \sim \mathrm{2-3}\Gyr$ \citep[e.g.,][]{Bigiel.etal.2008}. In the Milky Way this time scale is somewhat shorter: $t_{\rm dep, MW} \sim 10^9\Msun/1.5\Msunyr \sim 0.7 \Gyr$.

In our simulation without any tuning we obtain the global depletion time of $\sim 1 \Gyr$, which is similar to that of the Milky Way. However, given that in our simulation local consumption time exhibits wide variation (Figure~\ref{fig:1:disk}f), this global depletion time is a result of averaging regions, that form stars relatively fast (with $\tdep \sim 100 \Myr$), with gas, that forms stars slowly (with $\tdep > 3 \Gyr$). Observationally, such spatial variation of $\tdep$ can be studied in high resolution extragalactic SFR surveys \citep[e.g.,][]{Rebolledo.etal.2015, Lewis.etal.2015}.

Figure~\ref{fig:4:KS} compares the KS relation for our simulated galaxy with the relations observed for 7 nearby spirals in patches of 750 pc \citep{Bigiel.etal.2008} and measured in the Milky Way in radial annuli \citep{Misiriotis.etal.2006}. In our simulation we integrate the SFR and gas density perpendicular to the disk plane and then average them in patches of $1\times1\kpc$. The star formation is temporally averaged over $30\Myr$ to match the averaging scales corresponding to star formation indicators typically used to estimate SFR in observations \citep[e.g., see Table 1 in][]{Kennicutt.Evans.2012}. The figure shows that our simulation predicts the KS relation in broad agreement with observations. {\it We stress that this agreement is achieved without any ad hoc assumptions about star formation efficiency or threshold:} our results are based on direct application of the star formation efficiency prescription, calibrated on simulations of turbulent GMCs, in a modelling of global disk evolution. 

At $\Sgas \sim 10 \Msunpc2$ our simulation predicts the relation that is somewhat steeper than observed. This may be due to the specifics of the disk structure and the dynamics of the galaxy we model. For example, the KS relation in individual galaxies studied by \citet{Bigiel.etal.2008} exhibits significant variation in normalization and slope. For instance, NGC628 and NGC3184 also have a steep relation between SFR and gas surface densities, similar to the distribution in our simulation. Variation of the KS relation with the disk properties thus requires further investigation. 

Another factor affecting predicted KS relation at low surface densities is the averaging time scale of star formation rate. For example, adopting the averaging time scale of $100\Myr$ instead of $30\Myr$ results in a noticeably flatter KS relation at low $\Sgas$. Note also that the absence of star-forming regions with $\Ssfr < 10^{-4} \Msunyrkpc2$ in Figure~\ref{fig:4:KS} is an artefact of numerical resolution and is set by the minimal mass of stellar particles we form in our simulation ($10^4 \Msun$). 

Finally, at the highest surface densities, the KS relation predicted in our simulation is shaped by the star formation in high-density clumps. Although such regions are absent in the Milky Way itself, they are often observed in other galaxies. For instance, the clump-driven star formation regime may have implications for starburst galaxies, which have surface densities and SFRs comparable or even higher than the highest values in our simulations \citep[e.g.,][]{Kennicutt.1998,Kennicutt.Evans.2012}. This is because starburst galaxies tend to have higher gas fractions and gas surface densities and are, therefore, more unstable. As a result, they develop more self-gravitating clumps that have higher $\epsff$ and shift the global SFR towards higher values for a given surface density. 

Our convergence tests indicate that with our implementation of feedback the distribution of SFRs averaged on $1\kpc$ scales does slightly shift with resolution. Such dependence indicates that the feedback in our simulation is not efficient enough to self-regulate the star formation and erase dependence of its large-scale properties on the assumptions about local SFE that do depend on resolution. This behavior agrees with findings of \citep{Agertz.Kravtsov.2015} who showed that high local SFE is required to reproduce the observed properties of galaxies via feedback self-regulation. As they also showed $\epsff\sim1\%$, that is the average SFE in our disk, is not enough to make the feedback efficient. However, the actual values of SFE in our simulation may be somewhat low due to overestimation of $\vrms$ in star-forming regions resulting from the poorly resolved onset of turbulent cascade. We will present a detailed discussion of the resolution tests and performance of our model depending on feedback recipes in a follow-up paper.

\section{Discussion}
\label{sec:discussion}

The results presented in the previous section show that our turbulent model for star formation predicts the wide variation of SFE from $\epsff < 0.1\%$ to $\sim 10\%$. The predicted distribution of SFE agrees with SFRs observed on a wide range of scales: from the scale of individual star-forming regions ($\sim 40$ pc) to $\sim\kpc$ scales and the scales of entire galaxies. This agreement is non-trivial because on kpc scales our model predictions are determined by the small-scale spatial distribution of density and turbulent energy shaped by galactic evolution. Moreover, on the scale of individual star-forming regions (cells in simulations) the star formation rate is not tuned but is estimated from the local gas density and turbulent energy using predictions of GMC-scale simulations (see Section~\ref{sec:method:SFR}). Once calibrated to reproduce the results of such simulations, parameters of both the SGS turbulence model and the prescription for star formation remain fixed in our galactic disk simulations. Remarkably, the model predicts SFRs in agreement with observations without any additional tweaking of these parameters.

Although the agreement with the observed KS relation can also be achieved in simulations that adopt universal $\epsff$ recipes, such recipes require ad hoc assumptions about a value of $\epsff$ and criteria for star formation. 

The obtained variation of SFE may have important implications for understanding self-regulation of star formation during galaxy evolution. A number of studies showed that in order to reproduce cosmic star formation histories and resulting morphology of Milky Way-sized galaxies simulations require high local SFE \citep[e.g.,][and references therein]{Hopkins.etal.2014,Agertz.Kravtsov.2015}. The low global SFE in such simulations is then achieved as a result of the star formation self-regulation by effective feedback. The star formation model in our disk simulation, on the other hand, explicitly predicts that in galaxies with average $\epsff\sim1\%$ certain regions may form stars very efficiently (with $\epsff\sim10\%$), regardless of the global SFR. 

The predicted localized efficient star formation also has implication for outflow driving. The concentration of massive stars in regions of high $\epsff$ means that SNe can explode there in a correlated manner, which can facilitate launching of fountain flows or even galactic winds \citep[e.g.,][]{Governato.etal.2010}. We have checked that a re-simulation of our disk with the same feedback model but $\epsff$ fixed at the value of $1\%$ results in no gas outflows, in contrast to our fiducial simulation, which sustains prominent fountain outflows. We will present a detailed comparison of these simulations in a subsequent paper. 

The second important feature of our model is the pronounced physical density threshold for star formation at $n_{\rm th} \sim 10\cc$. Such density threshold is often set by hand in galaxy formation modelling. In our simulation, however, this threshold arises from the rapid drop in temperature as density increases beyond $n_{\rm th}$. Such threshold is quite similar to the effective thresholds in H$_2$-based models of star formation in which local SFE is modulated by the molecular gas mass fraction $\fH2$: $\rhosfr = \epsff \fH2 \rho/\tff$ \citep{Robertson.Kravtsov.2008,Gnedin.etal.2009}. In particular, the molecular fraction correlates with the cold gas abundance and, therefore, the threshold in our model corresponds to the density at which $\fH2$ rapidly increases. We thus also expect that $n_{\rm th}$ in our model should depend on gas metallicity similarly to the threshold in the $\fH2$-based models if gas thermodynamics is modelled properly to capture dependence of the net cooling function on gas metallicity. 

Modelling of the star formation density threshold can be potentially improved if we take into account effects of gas clumpiness on its net cooling rate: $\Lambda_{\rm cool}\propto C$, where $C \equiv \langle \rho^2 \rangle / \langle \rho \rangle^2 \geq 1$ is a clumping factor and brackets denote averaging over a certain spatial region. Galaxy formation simulations almost always assume $C=1$ on the unresolved scales, but actual dense ISM can be quite clumpy in regions where turbulence is supersonic. Local clumping factor can be derived from the shape of the underlying subgrid density PDF, that can be estimated, for instance, with the aid of the SGS turbulence model from the local parameters, such as effective Mach number. The clumping factor could then be accounted for in the calculation of the net cooling rate $\Lambda_{\rm cool}$. Overall, the star formation threshold should shift to lower values of density for $C>1$.

The importance of turbulence for modelling star formation in galaxy formation simulations has been already recognized. Specifically, \citet{Hopkins.etal.2013} developed a model for the star formation threshold, in which star formation is allowed only in self-gravitating gas, $\avir < 1$, but with a constant efficiency of $\epsff = 100\%$. Even though their model also predicts significant localization of star formation, it substantially differs from our model both technically and conceptually. 

From the technical point of view, our SGS model provides a more appropriate way to track local $\vrms$ than an estimate based on the local velocity gradients on the scale of resolution. Due to substantial effects of numerical viscosity on these scales such estimate is not accurate.

More importantly, in our model we vary $\epsff$ continuously with $\avir$, as predicted by the \citet{Padoan.etal.2012} model, while the prescription of \citet{Hopkins.etal.2013} adopts a fixed $\epsff = 100\%$ for $\avir < 1$ and $\epsff = 0$ otherwise. For comparison, the \citet{Padoan.etal.2012} model predicts $\epsff \approx 26\%$ for $\avir \approx 1$ and $\epsff$ reaches $>99\%$ only at $\avir\lesssim 0.1$. Moreover, in our model star formation can proceed in gravitationally unbound regions. In the turbulence-driven star formation, a given region may be globally unbound, but can contain local bound star-forming regions created by the turbulent cascade on small scales. For example, at $\avir\approx 10$ the \citet{Padoan.etal.2012} model predicts $\epsff\approx 1\%$, which is a healthy efficiency estimated for many GMCs \citep[e.g.,][]{Krumholz.etal.2012}. Thus, the sharp threshold in $\avir$ for $\epsff$ is not warranted.

Turbulent models of star formation with $\epsff$ continuously varying with $\avir$ were studied by other authors as well. In particular, \citet{Braun.etal.2014} examined a star formation prescription based on the model of \citet{Padoan.Nordlund.2011} coupled with an SGS turbulence model in isolated disk simulations. Their subgrid model also included a prescription for multiphase ISM \citep{Braun.Schmidt.2012}, and turbulent velocities were rescaled to the scale of cold self-gravitating clumps within this subgrid medium. Our disk models and star formation implementations are sufficiently different, which complicates a direct comparison of our results. We only note that although \citet{Braun.etal.2014} also found gas depletion time variation across the galaxy, in their case this variation was mostly due to the variation of $\fH2$, as the SF prescription based on the \citet{Padoan.Nordlund.2011} results predicted $\epsff\sim10\%$ in their star-forming regions with only small scatter. The variation of the depletion time in our simulations is due to the wide variation of local $\epsff$, which, in turn, is caused by the scatter in the virial parameter $\avir$ that is dominated by the variation of turbulent velocities.

This origin of variation of SFE and gas depletion in our model is more in line with the models studied more recently by \citet{Braun.Schmidt.2015}. These authors used a series of disk simulations similar to those in \citet{Braun.etal.2014}, but examined several star formation prescriptions based on the local turbulent properties predicted by the SGS model, including the model of \citet[][the ``PHN'' model]{Padoan.etal.2012}, which we use in our work. 

Although the overall level of turbulence predicted by the SGS model in their disk is in qualitative agreement with our results, \citet{Braun.Schmidt.2015} found that the PHN model predicts values of SFE that are systematically too low: $\epsff \lesssim 0.1\%$. According to Equation~(\ref{eq:SFRlaw}), such low $\epsff$ should arise in regions with virial parameter: $\avir \gtrsim 20$, which is significantly larger than the values estimated for the observed GMCs \citep[e.g.,][]{Bolatto.etal.2008,Dobbs.etal.2011}. We believe that the origin of this discrepancy is in the subgrid model of multiphase gas distribution used by \citet{Braun.Schmidt.2015}. In this model, the size of cold clouds is set by the condition of $\avir=1$, but at the same time, turbulent velocities are rescaled from the cell size to the cloud size assuming a turbulent cascade scaling which results in $\avir \gtrsim 20$. This indicates that their model is not internally consistent.

Regardless of the difference in actual values of $\epsff$, the results of \citet{Braun.Schmidt.2015} are qualitatively consistent with our main finding that turbulence based star formation prescription predicts a wide variation of $\epsff$.

\section{Summary and conclusions}
\label{sec:summary}

In this paper we have presented a star formation prescription, in which local star formation efficiency is a function of local gas density and turbulent velocity dispersion, as predicted by direct simulations of star formation in turbulent molecular clouds. We have tested the model using simulations of an isolated Milky Way-sized disk galaxy that included a self-consistent treatment of turbulence on unresolved scales. Our main results and conclusions can be summarized as follows. 

\begin{itemize}

\item[1.] We find that in our simulations the shear-improved subgrid turbulence model of \citet{Schmidt.etal.2014} predicts turbulent rms velocities on the resolution scale of our simulations ($40\pc$) consistent with the observed velocity dispersions in Milky Way GMCs and nearby galaxies. 

\item[2.] The level of turbulence in dense, star-forming gas is largely determined by the enhancement of weak turbulence in warm diffuse gas by compression in spiral density waves. As a result, the turbulent velocities in the cold gas correlate with the local compression rate. Turbulence in the warm gas in between spiral arms, on the other hand, is maintained in equilibrium between the cascade from the resolved scales, sourcing by SNe and viscous dissipation into heat. 

\item[3.] The star formation model, in which efficiency scales exponentially with the ratio of local free-fall and turbulent crossing time, as found in the GMC simulations of \citet{Padoan.etal.2012}, predicts a wide variation of star formation efficiency, from $\epsff \lesssim 0.1\%$ to $\sim 10\%$, and local gas depletion time, from $\tdep \sim 100 \Myr$ to $\tdep > 3 \Gyr$. 

\item[4.] We find that the predicted distribution of $\epsff$ does average to the observed global depletion time $\tdep \sim 1 \Gyr$, typical for Milky Way-sized galaxies. Moreover, the resulting distribution of star formation rates in individual star-forming regions broadly agrees with observations of star formation in the Milky Way and nearby galaxies. This agreement is achieved without any tuning of our star formation model parameters.

\item[5.] In addition, our SF prescription naturally predicts a physical threshold for star formation at the density, at which the thermal pressure of warm diffuse gas starts to dominate over turbulent pressure. This threshold density is expected to vary with the gas metallicity in a manner similar to the molecular fraction, although it does not depend on the molecular gas physics explicitly.  

\end{itemize}

All in all, we conclude that the presented implementation of star formation based on turbulence produces realistic star formation efficiencies and rates when applied in a galaxy-scale simulation. The lack of free parameters and the fact that this prescription relies on direct GMC-scale simulations put star formation modelling within this framework on a much firmer footing compared to standard recipes. This makes the described SF prescription promising for use in cosmological galaxy formation simulations, which we will pursue and present in a subsequent study.

\acknowledgments

We are deeply grateful to Alexei Kritsuk for valuable discussions and for providing us with the initial conditions for developed turbulence, that were used for testing our SGS turbulence model. This work was supported by a NASA ATP grant NNH12ZDA001N, NSF grant AST-1412107, and  by the Kavli Institute for Cosmological Physics at the University of Chicago through grant PHY-1125897 and an endowment from the Kavli Foundation and its founder Fred Kavli. The simulations presented in this paper have been carried out using the Midway cluster at the University of Chicago Research Computing Center, which we acknowledge for support.

\bibliographystyle{aasjournal}
\bibliography{}

\end{document}